\documentclass[aps,pre,twocolumn,amsmath,amssymb,letterpaper]{revtex4-1}
\usepackage{graphicx}

\allowdisplaybreaks[1]

\begin{document}
\title{Helical edge resistance introduced by charge puddles}
\author{Jukka I. V\"{a}yrynen}
\affiliation{Department of Physics, Yale University, New Haven, CT 06520, USA}
\author{Moshe Goldstein}
\affiliation{Department of Physics, Yale University, New Haven, CT 06520, USA}
\author{Leonid I. Glazman}
\affiliation{Department of Physics, Yale University, New Haven, CT 06520, USA}
\date{\today}
\begin{abstract}
We study the influence of electron puddles created by doping of a 2D topological insulator on its helical edge conductance.
A single puddle is modeled by a quantum dot tunnel-coupled to the helical edge. It may lead to significant inelastic backscattering within the edge because of the long electron dwelling time in the dot. We find the resulting correction to the perfect edge conductance. Generalizing to multiple puddles, we assess the dependence of the helical edge resistance on temperature and doping level, and compare it with recent experimental data.
\end{abstract}
\maketitle

The realization that a boundary separating a topologically-nontrivial insulator from a conventional one should carry delocalized electron states \cite{kane05a,kane05b} has led to the prediction of such states in concrete materials and their experimental observation
\cite{bernevig06,kane_review,qi_review}.
%Their existence %of these surface states
%(or edge states, in the case of 2D insulators)
%was confirmed experimentally by a variety of techniques \cite{kane_review,qi_review}.
One of the stunning theoretical predictions is that in 2D the zero-temperature electron transport along an edge is reflectionless, as long as time-reversal symmetry is not broken, which should lead to the quantization of the edge conductance \cite{kane05b}.

Experiments with HgTe quantum wells
of the appropriate thickness
confirmed the existence of highly-conducting channels in a nominally insulating state of a heterostructure \cite{konig07,konig08,roth09}. The Fermi energy $E_F$ in a heterostructure was tuned by a gate to reside in the gap between the valence and conduction bands. The values of conductance $G$ measured under these conditions were indeed close to the predicted quantized value $G_0=e^2/h$ per edge, but only for small $\sim 1\times 1\,\mu{\rm m}^2$ samples. Deviations $\Delta G \equiv G_0 - G$ towards lower conductance values were clearly seen in larger samples
%for samples of lengths $L\gtrsim 1\,\mu{\rm m}$
\cite{konig07,konig08,roth09,gusev11,konig12,nowack12}.

In short samples, $\Delta G$ fluctuated with gate voltage. The temperature dependence of $G$ has not been systematically measured yet, but the existing data indicate it to be rather weak. %relatively weak evolution of the conductance with temperature.
These observations should be contrasted with the theoretical predictions of a strong temperature dependence of electron inelastic backscattering rate, with %the
a characteristic scale %for temperature
set by the band gap $E_G$. Depending on the model, $\Delta G$ scales as $\propto (T/E_G)^6$ or $(T/E_G)^4$, unless time-reversal symmetry is broken~\cite{kane05b,xu06,schmidt12,lezmy12}.
%In the absence of an external magnetic field that would require a
Spontaneous symmetry breaking
%, which
is improbable for %in the case of
%relatively
weak electron-electron interaction (noting the high dielectric constant, $\kappa\approx13$ \cite{qi_review}) and in the absence of a high density of magnetic impurities. Luttinger liquid effects~\cite{kane05b,xu06,strom10,lezmy12} are also suppressed in short samples.
%(hereinafter we dispense with the Luttinger liquid effects, cf., e.g., \cite{kane05b,xu06,strom10,lezmy12} as the samples were relatively short).

\begin{figure}[b]
\includegraphics[width=7cm,height=!]{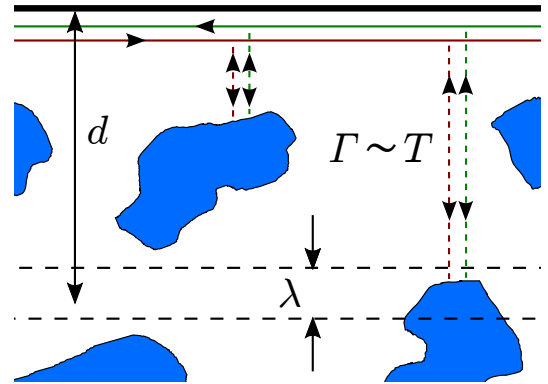}
\caption{\label{fig:system}
(Color online)
Electrons moving along a helical edge tunnel in and out of puddles created by the inhomogeneous charge distribution in the heterostructure. In the puddles electrons may undergo inelastic backscattering. The main contribution %to backscattering
comes from puddles whose distance $d$ from the edge is within a strip where the resulting level width $\Gamma \sim T$, cf.\ Eqs.~(\ref{eq:peak-av1})--(\ref{eq:peak-av2}). The strip width is the tunneling length $\lambda = v/E_G$. Summation over the puddles yields the average resistivity, Eq.~(\ref{eq:G-av3}).
}
\end{figure}

The existing theory considers inelastic electron backscattering by either
%in two limits: the first one corresponds to
uniform interactions along an edge \cite{kane05b,strom10,budich12},
%and the second one refers to interactions
or at isolated points %along an edge
\cite{kane05b,xu06,footnote_kondo,schmidt12}. Helical edges formed in a semiconductor heterostructure are likely to deviate considerably from either limit. The structures are
%grown with some level of doping
doped~\cite{konig07,konig08,roth09,konig12,nowack12}; the presence of charged donors and acceptors results in a non-uniform potential landscape for electrons. These inhomogeneities are not point-like because of the long-range of the Coulomb potential.
%Further complication comes from the fact that
Moreover, the topologically non-trivial insulators are in fact narrow-gap semiconductors with a typical  gap of only $E_G\simeq 10\, \text{meV}$ for HgTe quantum wells~\cite{konig07,konig08,roth09,konig12,nowack12,novik05}. To place $E_F$ inside the band gap, an appropriate gate voltage is applied, so that the gate charge balances out the uncompensated donor ($n_d$) or acceptor ($n_a$) charge density.  The joint effect of the gate and ionized dopant atoms may lead to the formation of electron and hole puddles in the quantum well, cf.\ Fig.~\ref{fig:system},
%, as depicted in Fig.~\ref{fig:system},
similar to the known phenomenon in compensated bulk semiconductors \cite{efros}.
%Indeed, in
In the existing measurements doping varied from $n_d\sim 3.5\cdot 10^{11}\,\text{cm}^{-2}$ to $n_a\sim 5\cdot 10^{10}\,\text{cm}^{-2}$ \cite{konig07,konig08,roth09,konig12,nowack12}, and
the results indeed seem to indicate that a lower doping level improves the quality of %helical
the edge conductance quantization.
Furthermore, the uncovered strong sensitivity of the edge conductance to the potential of a scanning probe \cite{konig12} may imply the presence of puddles, i.e., spontaneously formed quantum dots, in the vicinity of the edge.

In this Letter we elucidate the role of tunneling between an edge and a quantum dot on the
%conduction of an
edge conductance.
%Remarkably, the topological protection ensures that the
Elastic processes involving electron dwelling in the dot do not lead to any backscattering. However, dwelling enhances the inelastic backscattering
%rate
by increasing the time electrons interact with each other. % in a scattering event.
At temperatures
%In a realistic regime of temperature
$T < \delta$, the dwelling-time effect makes the conductance correction strongly dependent on the position of the Fermi level $E_F$ with respect to the dot energy levels, and on the tunneling widths $\Gamma$ of these levels ($\delta\ll E_G$ is the mean level spacing in the dot). At a given temperature $T$, the tallest peaks $\Delta G^\text{peak} \propto (T/\delta)^2$ in $\Delta G (E_F)$ are produced by levels with $\Gamma\sim T$, see Eq.~(\ref{eq:peak-av2}) and Fig.~\ref{fig:system}. Such peaks in $\Delta G(E_F)$ are of widths $\sim T$, and the ``peak-to-valley'' ratio is $\sim (\delta/T)^6$.

Dots, or puddles of charge carriers in a quantum well, are formed accidentally by fluctuations in the donor density~\cite{gergel78,efros93}. We establish a crossover value $n_0$ of $n_d$ below which puddles are rare. At $n_d\ll n_0$ the density of puddles, $n_p$, is exponentially small in $n_0/n_d$. In short samples of length $L\lesssim n_p^{-1/2}$ only a few puddles are in the vicinity of the edge, resulting in mesoscopic fluctuations of $G$ with the gate voltage. This model agrees with the results of scanning-gate experiments \cite{konig12} and could explain the variations of $G$ with the back gate voltage in earlier experiments~\cite{konig07,konig08,roth09}, if the condition $n_d\lesssim n_0$ would hold there. For longer samples, $L\gg n_p^{-1/2}$, many puddles couple effectively to the edge. That leads to edge resistance, $R\propto n_p L (T/\delta)^3$, which varies smoothly with the gate voltage and possibly greatly exceeds the quantized value $h/e^2$.  At the same time, the ``bulk'' hopping conductivity, which is proportional to factors exponentially small in $n_p^{-1/2}$ and $T/\delta$, may still remain negligible. In this case, current would flow along the edges, despite edge resistance being high compared to $h/e^2$, as observed in Ref.~\cite{nowack12}. The model would also explain the earlier measurements~\cite{konig07,konig08,roth09,gusev11} on larger samples, if the condition $n_d\lesssim n_0$ would be satisfied [our crude estimate of $n_0$, Eq.~(\ref{eq:n0}), turns out to be too low for that].

We start by considering a helical edge coupled to a single quantum dot via a point contact. In the absence of interactions, the corresponding Hamiltonian takes form:
\begin{align}
\hat{H}_0 = & -iv_F \sum_{\gamma}\gamma\int dx
\psi^\dagger_\gamma(x)\partial_x\psi^{\phantom\dagger}_\gamma(x)
+\sum_{n\gamma}\varepsilon_{n}c_{n\gamma}^{\dagger}c_{n\gamma}
\nonumber\\
& + \sum_{n,\gamma}t_{n}c_{n\gamma}^{\dagger}
\psi^{\phantom\dagger}_{\gamma}(0)
+ \text{H.c.}.
\label{eq:FreeHamiltonian}
\end{align}
Here $v_F$ is the helical edge velocity, $\gamma=\pm 1 \equiv R,L$ labels the right- and left-movers, respectively, and $n$ labels the discrete energy levels in the dot, measured from $E_F$. The dot is coupled to the edge at $x=0$ by a set of tunneling amplitudes $t_n$. The Kramers degeneracy of each discrete energy level $n$ gave us the freedom to pick the corresponding eigenfunctions $|n\gamma\rangle$ in such a way that the left- and right-movers are coupled to two different components of each doublet. %One can see from the structure of the Hamiltonian~(\ref{eq:FreeHamiltonian}) that
There is thus no backscattering in the free-electron problem. Interaction in the dot,
\begin{equation}
\hat{U}= \frac{1}{2}\!\!
\sum_{n_{i}, \gamma_{i}}
%\sum_{n_{1} \dots n_{4}}\!\!\sum_{\gamma_{1}\dots \gamma_{4}}
U_{n_{1}\gamma_{1}n_{2}\gamma_{2};n_{3}\gamma_{3}n_{4}\gamma_{4}}
c_{n_{1}\gamma_{1}}^{\dagger}c_{n_{2}\gamma_{2}}^{\dagger}
c^{\phantom\dagger}_{n_{4}\gamma_{4}}
c^{\phantom\dagger}_{n_{3}\gamma_{3}},
%\nonumber\\
\end{equation}
may lead to inelastic backscattering (hereinafter, we assume $\hat{U}$ respects time-reversal symmetry).

The inelastic backscattering reduces the steady-state current $I=I_0-\Delta I$ from its ideal value $I_0 = G_0 V$ by
\begin{align}
\Delta I
 &= e\sum_{\gamma_i}
 \Delta N_{\gamma_{1}\gamma_{2};\gamma_{3}\gamma_{4}} \!\!
 \int dE_{1}dE_{2}dE_{3}dE_{4}
 \nonumber\\
 &
 \times S_{\gamma_{1}\gamma_{2};\gamma_{3}\gamma_{4}}(E_{1},E_{2};E_{3},E_{4})
 \delta(E_1+E_2-E_3-E_4)
 \nonumber\\
 &\times \big[ \tilde{f}_{\gamma_1}(E_1) \tilde{f}_{\gamma_2}(E_2) (1-\tilde{f}_{\gamma_3}(E_3))(1-\tilde{f}_{\gamma_4}(E_4))
 \nonumber\\ &
 \quad
 - \tilde{f}_{\gamma_3}(E_3)\tilde{f}_{\gamma_4}(E_4) (1{-}\tilde{f}_{\gamma_1}(E_1)) (1{-}\tilde{f}_{\gamma_2}(E_2))
 \big]\,. \label{eq:deltaI}
\end{align}
Here $V$ is the source-drain voltage,
$\Delta N_{\gamma_{1}\gamma_{2};\gamma_{3}\gamma_{4}} = (\gamma_{1}+\gamma_{2}-\gamma_{3}-\gamma_{4})/2$ counts the net number of right-movers scattered into left-movers,
$\tilde{f}_{\gamma_i}(E) = 1/[e^{(E+\gamma_i eV/2)/T}\!+1]$ is the Fermi function shifted by $\pm eV/2$, and
$S_{\gamma_{1}\gamma_{2};\gamma_{3}\gamma_{4}}(E_{1},E_{2};E_{3},E_{4})$
is the cross section for the two-electron scattering process
$|E_1 \gamma_1, E_2 \gamma_2\rangle \rightarrow |E_3 \gamma_3, E_4 \gamma_4\rangle$
between exact left- and right-propagating eigenstates of the Hamiltonian~(\ref{eq:FreeHamiltonian}). In general, $S$ allows for backscattering of one ($RR \to LR$) or two ($RR \to LL$) electrons. There are two respective contributions, $\Delta G_1$ and $\Delta G_2$, to the conductance $G=G_0-\Delta G_1-\Delta G_2$.

In the Born approximation the cross section is
\begin{align}
&S_{\gamma_{1}\gamma_{2};\gamma_{3}\gamma_{4}}(E_{1},E_{2};E_{3},E_{4})
=\frac{2}{\pi^{3}}
\sum_{m_i, n_i}
\left[\prod_{i=1}^4 \text{Im} \mathcal{G}^R_{n_i m_i} (E_i) \right]
\nonumber \\ &
\qquad
\times
U_{m_{1}\gamma_{1}m_{2}\gamma_{2};m_{3}\gamma_{3}m_{4}\gamma_{4}}^{*}
U_{n_{1}\gamma_{1}n_{2}\gamma_{2};n_{3}\gamma_{3}n_{4}\gamma_{4}}\,.
\label{eq:S}
\end{align}
Here $\mathcal{G}^R_{n_1 n_2}(E)$ is the noninteracting retarded Green function of an electron in the dot. All interaction matrix elements
%of Eq.~(\ref{eq:InteractionHamiltonian})
must be small compared to $\Gamma$ to allow the perturbative treatment at arbitrary position of the Fermi level with respect to the dot levels. This condition is
more easily satisfied %easier
for the off-diagonal matrix elements~\cite{aleiner02} entering explicitly in Eq.~(\ref{eq:S}), than for the diagonal ones
$U_{n_1\gamma_1 n_2\gamma_2;n_1\gamma_1 n_2\gamma_2} \sim E_C$. The introduced charging energy $E_C$ is small, $E_C\ll\delta$, if the spacer between the quantum well and gate is thinner than the Debye radius for electrons in the well. In the opposite case of $E_C\gtrsim\delta$,  Coulomb blockade may develop. We first treat
%Treating
the entire interaction perturbatively,
%we dispense with it, but will
and later point out
%later
how Coulomb blockade modifies
%our
the results. We will also see that backscattering is dominated by puddles with $\Gamma \sim T$; thus Kondo correlations~\cite{footnote_kondo} setting in at the much lower temperature $T_K \ll \Gamma$ can be ignored.

Using properties of the interaction matrix elements in Eq.~(\ref{eq:S}), it is straightforward to check that in the low-temperature limit $\Delta G_1\propto T^4$ and $\Delta G_2\propto T^6$, in agreement with Refs.~\cite{kane05b,xu06,schmidt12,lezmy12}. For a generic form of strong spin-orbit interaction in the dot, all interaction matrix elements in Eq.~(\ref{eq:S}) are of the same order~\cite{aleiner02}.
%(we assume the electron states in the dot are described by a Gaussian symplectic ensemble) \cite{aleiner02}.
In this case, $\Delta G_2/\Delta G_1\ll 1$ if $T\ll\delta$. The proportionality coefficient of the temperature dependence $\Delta G_1\propto T^4$ is a function of the dot parameters; in the case of weak tunneling it peaks every time a level crosses the Fermi energy.

Weak tunneling corresponds to small elastic tunnel widths $\Gamma_n=|t_n|^2/(2v_F)$ of the levels, $\Gamma_n\ll |\varepsilon_n-\varepsilon_{n\pm 1}|$. %In that limit,
Then the leading-order approximations for the diagonal and off-diagonal ($n_1\neq n_2$)
matrix elements of $\hat{\mathcal{G}}^R(E)$ read $\mathcal{G}^R_{n n}(E) = (E - \varepsilon_{n} + i\Gamma_{n})^{-1}$ and $\mathcal{G}^R_{n_1 n_2}(E) =-i\sqrt{\Gamma_{n_1}\Gamma_{n_2}} [(E - \varepsilon_{n_1} + i\Gamma_{n_1})(E - \varepsilon_{n_2} + i\Gamma_{n_2})]^{-1}$, respectively. %($n_1\neq n_2$).
Using this simplification in Eq.~(\ref{eq:S}) we find
\begin{equation}
  \label{eq:dg1_peak}
 \frac{\Delta G_1^\text{peak}}{G_0}
  =
  \frac{2^7 \pi}{15} \left(\frac{T}{\Gamma_1}\right)^4
  \left|
  \sum_{n \ne 1}
  \sqrt{\frac{\Gamma_n}{\Gamma_1}}\cdot\frac{U_{1L 1R; 1R nR}}{\varepsilon_n}
  \right|^2
\end{equation}
for the peak in $\Delta G_1$ corresponding to the level $\varepsilon_1$ crossing the Fermi level ($\varepsilon_1=0$). The peak height and its width, $|\varepsilon_1-E_F|\sim\Gamma_1$, display mesoscopic fluctuations; Eq.~(\ref{eq:dg1_peak}) is applicable at $T\ll\Gamma_1$. The peak value $\Delta G_1^\text{peak}$ grows with temperature till $T$ reaches a value $T\sim\Gamma_1$. At higher temperatures, some of the incoming electrons with energies $|E|\lesssim T$ which contribute to $\Delta G_1$ are off resonance. This leads to a decreasing $T$-dependence of $\Delta G_1^\text{peak}$ at $T\gtrsim\Gamma_1$,
\begin{equation}
  \label{eq:dg1_peak_t}
  \frac{\Delta G_1^\text{peak}}{G_0}
  = \frac{\Gamma_1}{T}
  \left|
  \sum_{n \ne 1}
  \sqrt{\frac{\Gamma_n}{\Gamma_1}}\cdot\frac{U_{1L 1R; 1R nR}}{\varepsilon_n}
  \right|^2\, ,
\end{equation}
and a peak width $|\varepsilon_1-E_F|\sim T$.

In a weakly-disordered dot the Thouless energy $E_T=g\delta\gg\delta$ ($g\gg 1$ is the dimensionless conductance within the dot). The disorder-averaged matrix elements $\langle U^2\rangle$ of interaction present in Eqs.~(\ref{eq:dg1_peak}) and (\ref{eq:dg1_peak_t})  can be evaluated using the standard diagrammatic techniques~\cite{aleiner02}. Further simplification %of the result
is possible for the screened Coulomb interaction, which is dominated by its universal zero-momentum component,
%: its zero-momentum component is universal, while components with $k\sim k_F$ are small (here $k_F$ is the Fermi momentum of electrons in the dot) yielding
leading to $\langle U^2\rangle \sim \delta^2/g^2$. Using this estimate in Eqs.~(\ref{eq:dg1_peak})--(\ref{eq:dg1_peak_t}) and dropping numerical factors, we arrive at the interpolation
\begin{equation}
\frac{\langle\Delta G_1^\text{peak}\rangle}{G_0}
\sim
\frac{1}{g^2}\frac{T^4}{\Gamma^4}\theta (\Gamma - T)
+
\frac{1}{g^2}\frac{\Gamma}{T}\theta (T-\Gamma)
\label{eq:peak-av1}
\end{equation}
for the typical peak conductance as a function of $T$ at small charging energy, $E_C\ll\max\{T,\Gamma\}$.
%. Equation~(\ref{eq:peak-av1}) assumes the charging energy is small, $E_C\ll\max\{T,\Gamma\}$.

The backscattering processes leading to Eqs.~(\ref{eq:dg1_peak}) and (\ref{eq:dg1_peak_t}) involve a sequence of virtual
states. Those with energy deficit $|\varepsilon_n|\neq 0$ are represented by the denominators in the sums over $n\neq 1$. One of the virtual states, however, has two electrons on level $n=1$ and brings a large factor $\sim 1/\Gamma^2$ to Eqs.~(\ref{eq:dg1_peak}) and (\ref{eq:dg1_peak_t}). It is replaced by $1/E_C^2$ in the presence of charging energy $E_C\gg \Gamma$. For the same reason, the cross section Eq.~(\ref{eq:S}) loses sensitivity to the energy $E_i$ of one of the two electrons involved. The corresponding integration range in Eq.~(\ref{eq:deltaI}) is restricted then by $T$ rather than $\Gamma$ at any $T/\Gamma$.
In the important (see below) case $E_C\sim\delta$,
the two modifications change Eq.~(\ref{eq:peak-av1}) by a factor $\sim(\Gamma/\delta)^2\cdot\max\{1,T/\Gamma\}$, leading to:
\begin{equation}
\frac{\langle\Delta G_1^\text{peak}\rangle}{G_0}
\sim
\frac{1}{g^2}\frac{T^4}{\Gamma^2\delta^2}\theta (\Gamma - T)
+
\frac{1}{g^2}\frac{\Gamma^2}{\delta^2}\theta (T-\Gamma)\,.
\label{eq:peak-av2}
\end{equation}
Backscattering in the``valley'' (Fermi level in between two subsequent dot levels) regime does not involve any low-energy virtual state and is not affected qualitatively by $E_C\sim\delta$. The corresponding estimate,
$\langle\Delta G_1^\text{valley}\rangle/G_0\sim T^4\Gamma^4/g^2\delta^8$ is smaller than the peak value Eq.~(\ref{eq:peak-av2}) by a factor $\sim(\Gamma^2/\delta^6)\cdot\max\{\Gamma^4,T^4\}$.

The main contribution to the backscattering correction averaged over the position of the Fermi level comes from the peak values, Eq.~(\ref{eq:peak-av2}), as $\langle\Delta G_1^\text{valley}\rangle/G_0$ is parametrically smaller. Accounting for the peak widths, $|\varepsilon_i-E_F|\sim\max\{\Gamma, T\}$, we find
\begin{equation}
\frac{\langle\Delta G_1^\text{av}\rangle}{G_0}
\sim
\frac{1}{g^2}\frac{T^4}{\Gamma\delta^3}\theta (\Gamma - T)
+
\frac{1}{g^2}\frac{\Gamma^2 T}{\delta^3}\theta (T-\Gamma)\,.
\label{eq:peak-av3}
\end{equation}

At higher temperatures the above
%(reminiscent of co-tunneling~\cite{averin89,averin90})
mechanism gives way to thermally-activated backscattering processes. Those originate only from the diagonal elements $\mathcal{G}^R_{nn}(E)$ in Eq.~(\ref{eq:S}).
Since this regime is probably not relevant for the interpretation of existing experiments (see below) we only sketch the results, deferring a detailed discussion~\cite{future}.
There are two types of activated contributions to $\Delta G$. The first one involves transitions within a pair of levels, $\{n_3,n_4\}=\{n_1,n_2\}$.
%, and takes the full advantage of the resonant tunneling processes.
The other one involves more levels, $\{n_3,n_4\}\neq\{n_1,n_2\}$, and gains importance at higher temperatures ($T\gg\delta$) due to the larger phase space available for transitions. At $T \lesssim \delta$ backscattering
is dominated by %the contribution coming from
the two levels closest to $E_F$, and $\Delta G \sim (\delta^2/g^2\Gamma T)\!\cdot\! \exp(-\varepsilon/T)$ with $\varepsilon\sim\delta$. Comparison with Eq.~(\ref{eq:peak-av3}) shows that activated backscattering exceeds $\langle\Delta G_1^\text{peak}\rangle$ at $T\gtrsim\delta/\ln(\delta/\Gamma)$. The distinction between peaks and valleys is lost at these temperatures, although $\Delta G$ does experience strong mesoscopic fluctuations at $T \sim \delta$ due to the randomness of the activation energy $\varepsilon$.

Now we turn to the typical experimental case~\cite{konig07,roth09,gusev11,konig08,novik05} of a doped, gate-controlled heterostructure. For definiteness, we will address the case of $n$-doped samples, assuming donors of average density $n_d$ are randomly distributed in a plane situated between the gate and quantum well; the distances of the donor plane and gate from the quantum well are $\ell_d$ and $\ell_g$, respectively. Random distribution of donors creates random potential $V({\mathbf r})$ for the charge carriers in the well. In the absence of carriers, the variance of the potential~\cite{gergel78} is $\langle V^2\rangle=V_0^2\ln \{\ell_g^2/[(2\ell_g-\ell_d)\ell_d]\}$ with $V_0=\sqrt{2\pi n_d} e^2/\kappa$ ($\kappa$ is the dielectric constant). At the point of full depletion (the gate charge density is $-en_d$) the probability of creation of electron and hole puddles depends on the ratio $E_G/(2\sqrt{\langle V^2\rangle})$. The relation $\sqrt{\langle V^2\rangle}=E_G/2$ defines a characteristic donor density,
\begin{equation}
n_0=
\frac{E_G^2\kappa^2}{8\pi e^4 \ln \{\ell_g^2/[(2\ell_g-\ell_d)\ell_d]\} }.
\label{eq:n0}
\end{equation}
The carrier puddles are small and rare if $n_d\ll n_0$; in the opposite limit ($n_d\gg n_0$), puddles are large and separated by thin depletion strips. In the following estimates we ignore the logarithmic factor in Eq.~(\ref{eq:n0}).

%We start with considering
In the limit $n_d\gg n_0$,
%. In this case, spatial
fluctuations of the
%conduction and valence
bands edges with respect to $E_F$ are large compared to $E_G/2$.
%, and spacings between puddles are small.
That
%also
allows us to use the linear approximation, $\varepsilon ({\mathbf k})=vk$,
%approximate
for the electron spectrum in the well~\cite{qi_review}.
% with a linear one, $\varepsilon ({\mathbf k})=vk$~\cite{qi_review}.
%To get an idea about the inter-puddle tunneling,
% rate between puddles,
%we note that
An electron
%wave function
penetrates over a length $\lambda\sim v/E_G$ into a p-n junction between the puddles. The junctions are formed by spatial fluctuations of the random potential $V({\mathbf r})$, and the typical width of the depletion region in a junction is
$E_G/|\nabla V({\mathbf r})|$. Tunneling is weak if $E_G/|\nabla V({\mathbf r})|\gtrsim \lambda$.
%; at the very least that condition has to be satisfied at $|{\mathbf r}_1-{\mathbf r}_2|\sim\lambda$.
To estimate the characteristic value of $|\nabla V({\mathbf r})|$,
%the gradient,
we use~\cite{footnote2} the correlation function $\langle|\nabla V({\mathbf r}_1)||\nabla V({\mathbf r}_2)|\rangle\sim V_0^2/|{\mathbf r}_1-{\mathbf r}_2|^2$ at $|{\mathbf r}_1-{\mathbf r}_2|\sim\lambda$ and
%  That allows us to
find that the weak-tunneling condition is $E_G/V_0\gtrsim 1$. It is {\sl not} satisfied at $n_d\gg n_0$;
%We conclude that
the p-n junctions are penetrable, leading to
%do not present exponentially large resistances for the charge transport, and expect the ratio
$\Gamma \gtrsim \delta$
%leading to the
and average bulk conductivity $\sigma_\text{bulk} \gtrsim e^2/h$.
%In this case, the single-puddle consideration is not applicable.
Recent analysis~\cite{fu12} suggests that a transition from the topological insulator to conductor state occurs at $\sigma_\text{bulk} \approx (1.4-2.5) e^2/h$. This makes the limit $n_d\gg n_0$ unfavorable for the helical edge conductance quantization at any temperature.

In the opposite limit, $n_d\ll n_0$, the puddles %of charge carriers in the quantum well
can be made rare by tuning the gate voltage.
In addition to $n_0$ there is another characteristic density scale, $1/a_B^2\approx 2\pi\alpha^4 n_0$, set by the effective Bohr radius $a_B=2\hbar v/\alpha E_G$ [we used $m^*=E_g/2v^2$ for the effective electron mass and $\alpha=e^2/(\kappa\hbar v)$ for the interaction parameter].
We restrict further consideration to the case $\alpha \ll 1$ and $n_0\gg n_d\gg 1/a_B^2$, where electrons form a continuous two-dimensional liquid at zero gate bias. Gate-induced depletion breaks the liquid into puddles, which become sparse once the electron chemical potential is shifted by $\gtrsim V_0$ below the conduction band edge~\cite{gergel78}. For
%weak interaction,
$\alpha \ll 1$ we may describe a puddle by
%may use
the Thomas-Fermi approximation.
%for the description of a puddle. %and repeating the same steps as in the $n_d\gg n_0$ case,
The puddle size $w$ is found by matching its number of electrons $N$ and electrochemical potential $N/(2 m^* w^2)$ to the typical impurity charge fluctuation $N \sim w n_d^{1/2}$ and potential fluctuation $V_0$, respectively.
%Then the puddle size $w$ is such that the number of electrons needed to compensate the typical impurity charge fluctuation $N \sim w n_d^{1/2}$ corresponds to a puddle
%%Fermi energy
%local electrochemical potential $N/(2 m^* w^2)\sim V_0$.
%%comparable to $V_0$.
This leads to $w\sim a_B$ and
%, and it is occupied by $N\sim a_B n_d^{1/2}$ electrons, and $\mu\sim e^2n_d^{1/2}/\kappa$; the level spacing is $\delta\sim\mu/N$, and the charging energy is $E_C \sim e^2/(\kappa w)$. %We cast $w$, $E_C$, $\delta$, and $g$ in the form
\begin{equation}
%w\sim\frac{1}{\sqrt{2\pi}\alpha^2n_0^{1/2}}\,;\,\,
E_C \sim \delta\sim \alpha^2 E_G\,;\,\, g \sim %\frac
{\left({n_d}/{n_0}\right)^{1/4}}/{(2\pi)^{1/4}\alpha}\,.
\label{eq:estimates2}
\end{equation}
Here we used estimates
%using the estimates
$\delta \sim V_0/N$, $g \sim \sqrt{N}$ \cite{aleiner02}, and assumed $w\lesssim\ell_g$, leading to
$E_C \sim e^2/(\kappa w)$.
%, and .
%to see that the estimates Eqs.~(\ref{eq:estimates1}) and (\ref{eq:estimates2}) match each other (up to unknown numerical factors) at $n_d\sim n_0$.
%In this regime
Puddles are located at rare strong fluctuations of the potential, and are thus far away from each other. Their density $n_p$ is estimated as the ratio of the total carrier density to the number $N\sim a_B n_d^{1/2}$ of electrons in a puddle. To find the former quantity we note that the distance to the gate $\ell_g$ serves as screening length for the potential fluctuations; hence we can divide the sample into roughly independent regions of size $\ell_g$.
A region becomes populated by carriers only if the local potential experiences an exponentially rare fluctuation exceeding $E_G/2$.
%Each such region will be populated by carriers only if the local potential experiences a fluctuation larger then $E_G/2$, and
The carrier number is such that they compensate for the fluctuation~\cite{gergel78}.
%the resulting number of carriers will be the one needed to compensate for that fluctuation~\cite{gergel78}.
This leads to
%The puddle density is then small,
$n_p \sim 1/(\ell_g a_B) (n_d/n_0)^{1/2} e^{-n_0/n_d}$.

The hopping conductivity facilitated by the puddles
%of the quantum well
is proportional to a product of two small parameters: the tunneling probability, exponential in $-(\lambda^2 n_p)^{-1/2}$; and the thermal activation probability, exponential in $-\delta/T$. The latter one remains %exponentially
small at $T\ll\delta$, even when approaching the crossover region $n_d \lesssim n_0$.
% \cite{efros}.
Under the same conditions, the rate of backscattering into the helical edge scales as a relatively low power of $T/\delta$, cf.\ Eqs.~(\ref{eq:peak-av2})--(\ref{eq:peak-av3}). For samples of length
$L\lesssim n_p^{-1/2}$ only a few puddles occur in the vicinity of the edge. That would make $\Delta G$ sensitive to a local probe potential, consistent with recent scanning gate measurements \cite{konig12}, and may also provide an explanation for the mesoscopic fluctuations of $\Delta G(E_F)$ in earlier measurements~\cite{konig07,konig08,roth09,konig12}. Eq.~(\ref{eq:peak-av2}) predicts that the largest peaks in $\Delta G(E_F)$ at a given temperature $T$ scale as $(T/\delta)^2$ and are produced by levels with $\Gamma\sim T$. For $L\gg n_p^{-1/2}$ contributions of many puddles add up incoherently, as scattering off each puddle is inelastic. The exponential dependence of the level widths $\Gamma \propto e^{-2 d/\lambda}$ on the distance $d$ between a puddle and the edge leads to a broad distribution of $\Gamma$. Hence, by Eq.~(\ref{eq:peak-av3}), backscattering will be dominated by puddles whose distance from the edge is such that $\Gamma \sim T$. Summing over puddles (see Fig.~\ref{fig:system}), we find that a long edge displays resistance $R=\rho_\text{edge}L$ with self-averaging resistivity
\begin{equation}
  \rho_\text{edge} \sim \frac{1}{G_0} \frac{1}{g^2} n_p \lambda \left( \frac{T}{\delta}\right)^3.
\label{eq:G-av3}
\end{equation}
While $\rho_\text{edge}\propto(T/\delta)^3$, the quantum well hopping conductivity $\sigma_\text{bulk}$ is exponentially small in $\delta/T$. Therefore, leakage into the bulk  at $n_d \lesssim n_0$ is insignificant for samples shorter than the exponentially-large ''leakage length'' $L^* = 1/(\sigma_\text{bulk}\rho_\text{edge})$, which may explain recent scanning SQUID results \cite{nowack12}.

Our findings thus match with observations provided that $n_d \lesssim n_0$ and $T \ll \delta$. To estimate $\delta$ for a HgTe/CdTe heterostructure, we use %Eqs.~(\ref{eq:estimates1}) and
Eq.~(\ref{eq:estimates2}) with $E_G = 10\text{meV}$  and $\alpha\approx 0.32$ (found with $\kappa\approx13$ and $v=5.5 \cdot 10^7 \text{cm}/\text{sec}$~\cite{qi_review}). We arrive at
%$\delta\approx 1-4\text{meV}$  at the crossover $n_d=n_0$;
$\delta\approx 1\text{meV}$,
comfortably above $k_B T$ in most experiments.
%at $n_d\lesssim n_0$;
%in most experiments $k_BT$ is comfortably below $\delta$.
For the crossover density, Eq.~(\ref{eq:n0}), we find $n_0 \approx 3 \cdot 10^{10} \text{cm}^{-2}$. The doping levels reported in Ref.~\cite{konig12} and in \cite{konig07,konig08,roth09} are, respectively, moderate
%normalized acceptor density
($n_a/n_0\sim 1$) and high ($n_d/n_0\sim 10$)  with respect to this value.
On the other hand, from the total resistance of long samples in Ref.~\cite{konig07} we deduce that $\sigma_\text{bulk} \lesssim 0.45 G_0$,
consistent with an insulating bulk~\cite{fu12}.
It may mean that our crude estimate of $n_0$ is off by a factor of $10$. The characteristic length $n_p^{-1/2}$, separating mesoscopic samples from the "self-averaging" ones, provides another check. % on theory.
The pre-exponential factor in it, $\sim 100\text{nm}$, is only $\sim 10$ times shorter than $1\mu{\rm m}$-long ``mesoscopic'' samples in Refs.~\cite{konig07,konig08,roth09}. That too may indicate that the samples doping was close to the true crossover value $n_0$.

To conclude, disorder in a doped heterostructure may lead to appreciable backscattering within a helical edge, while hopping conductivity in the quantum well remains negligible, which may explain some of the recent observations~\cite{konig07,konig08,roth09,konig12,nowack12}. The samples doping level $n_d$ apparently was close to the crossover value $n_0$ separating the regimes of low and high bulk conductivity, as opposed to our crude estimate, Eq.~(\ref{eq:n0}).
This complicates analysis of the of edge resistance dependence on $n_d$.
On the other hand, the robust qualitative features of the resistance $T$-dependence, in both the mesoscopic and the self-averaging regimes, Eqs.~(\ref{eq:peak-av2})--(\ref{eq:peak-av3}) and (\ref{eq:G-av3}), respectively, makes its detailed measurement very desirable.

%\begin{acknowledgments}
We thank D. Goldhaber-Gordon, K. Moler, and K. Nowack for stimulating discussions, and C. Varma for his request to write for the Journal Club for Condensed Matter Physics, which partially motivated this study. This work was supported by NSF DMR Grant No. 1206612, the Simons Foundation, and the Bikura (FIRST) program of the Israel Academy of Science.
%\end{acknowledgments}

\end{document}